\documentclass[aps,pre,superscriptaddress]{revtex4-1}
\usepackage{amsmath,amssymb}
\usepackage{graphics,graphicx,color}
\usepackage{dcolumn,bm}
\usepackage{tikz}

\definecolor{amethyst}{rgb}{0.6, 0.4, 0.8}

\def\[{\left[}
\def\]{\right]}
\def\({\left(}
\def\){\right)}
\def\be{\begin{equation}}
\def\ee{\end{equation}}
\def\bea{\begin{eqnarray}}
\def\eea{\end{eqnarray}}

\def\f{\mathbf{f}}

\def\nn{\nonumber}

\newcommand{\iisc}
{\affiliation{Centre for Condensed Matter Theory, Department of Physics, Indian Institute of Science, Bangalore 560012, India}}

\newcommand{\imsc}
{\affiliation{The Institute of Mathematical Sciences, Chennai 600113, India}}

\newcommand{\ncbs}
{\affiliation{Simons Centre for the Study of Living Machines, National Centre for Biological Sciences (TIFR), Bangalore 560065, India}}

\begin{document}
\title
{Extreme active matter at high densities
}

\author{Rituparno Mandal}%
\email[Email: ]{rituparno@ncbs.res.in}
\ncbs

\author{Pranab Jyoti Bhuyan}%
\email[Email: ]{pranab@physics.iisc.ernet.in}
\iisc

\author{Pinaki Chaudhuri}%
\email[Email: ]{pinakic@imsc.res.in}
\imsc

\author{Chandan Dasgupta}%
\email[Email: ]{cdgupta@physics.iisc.ernet.in}
\iisc

\author{Madan Rao}%
\email[Email: ]{madan@ncbs.res.in}
\ncbs

\begin{abstract}
Extreme active matter, an assembly of self-propelled particles with large persistence time $\tau_p$ and high P\'eclet number,
exhibits remarkable behaviour at high densities.  As $\tau_p\to 0$, the assembly
undergoes a gradual slowing down of density relaxations, as one reduces the active propulsion force $f$, until 
at the glass transition, the relaxation times diverge. 
In the other limit, $\tau_p \to \infty$, the fluid jams on lowering $f$, at a critical
threshold  $f^*(\infty)$, with stresses concentrated along force-chains. 
As one moves away from this jamming threshold,
the force-chains dynamically remodel, and the lifetime of the force-balanced configurations diverges as one approaches $f^*(\infty)$,
by tuning $\tau_p$. In between these limits, the approach to dynamical arrest at low $f$, goes through 
a phase characterised by intermittency in the kinetic energy. 
This intermittency is a consequence of long periods of jamming followed by bursts of plastic yielding associated with 
Eshelby deformations, akin to the response of dense amorphous solids to an externally imposed shear. 
The frequency of these plastic bursts increases as one moves towards the intermittent phase-fluid boundary, where
the correlated plastic events result in large scale vorticity and turbulence.
Dense extreme active matter brings together the physics of glass, jamming, plasticity and turbulence, in a new state of driven classical matter.
\end{abstract}

\pacs{61.20.Ja, 64.70.D-, 64.70.P-}
\maketitle

Active Matter, where each particle comprising the system is driven by an internal energy source and dissipates it in movement, 
constitutes a new class of driven non-equilibrium systems~\cite{activeRMP}.
{\it Extreme active matter}, where the magnitude of propulsion force is much higher than temperature and the direction of propulsion force persists over long times,
is an extreme realisation of activity. In this limit, active systems must show strong departures from equilibrium; this expectation
 is borne out in active Ornstein-Uhlenbeck particles (AOUPs) at low densities~\cite{fodor2016}, where steady states
with finite energy-current manifest when the persistence time is sufficiently large. 
Even so, one might suspect that at very high densities, these distinguishing effects of activity will be firmly suppressed
~\cite{bradley,angelini,kurchan,berthier,ni,marchetti1,marchetti2,mandal1,manning1,manning2,flenner,mandal2}. In this manuscript,
we see that extreme active matter at high densities is a fount of surprises, bringing
 together the physics of glass, jamming, plasticity and turbulence, in a new state of driven classical matter.

Our model dense extreme active matter is a dense assembly of self-propelled soft particles subject to a propulsion force of magnitude
$f$ and whose orientation persists over a time $\tau_p$.
Our main results are summarised in Fig.\,5:
(i) For small values of $\tau_p$, the assembly smoothly transforms from a fluid at high $f$ to a dynamically arrested glass
at low $f$. The phase boundary is well described by an active generalisation of RFOT theory (ARFOT)~\cite{ARFOT}, with an
``effective temperature'' that goes as $f^2\tau_p/(1+B\tau_p)$. 
However, we find that the mean kinetic energy has a different scaling behaviour with $\tau_p$, 
pointing to the feature that active systems should be characterised by many different ``temperatures''.
(ii) At intermediate values of $\tau_p$, the fluid abruptly transforms into an  {\it intermittent fluid} as $f$ is lowered to
$f^*(\tau_p)$, characterised by intermittency in the kinetic energy.
The intermittency increases as $f$ is reduced,  until at low enough $f$, the assembly  undergoes complete dynamical arrest.
(iii) This intermittency is a consequence of periods of jamming followed by bursts of plastic yielding.
 We identify  isolated plastic events with {\it Eshelby deformations}, akin to the response of dense amorphous solids to an externally imposed shear. 
(iv) As one approaches $f^*(\tau_p)$ from below, the plastic events become numerous and correlated in space-time, 
with an avalanche-like scale invariant statistics, that results in vorticity and turbulence, characterised by an
inverse cascade with a Kolmogorov exponent. 
(v)  The accumulated yielding over a time window involves the cooperative movement of a finite fraction of particles, and should
 manifest as a viscoelastic fluid at large time scales.
(vi) In the limit $\tau_p \to \infty$, the fluid reaches a jammed state on lowering $f$ to $f^*(\infty)$, with stresses concentrated along force-chains. 
As one moves away from this jamming threshold, by tuning $\tau_p$, the force-chains dynamically remodel, 
and the lifetime of the force-balanced configurations diverges as one approaches $f^*(\infty)$, with an exponent
$z \approx 0.71$.

The stochastic dynamics of a 2-dimensional assembly of interacting active brownian particles~\cite{abpreview}, 
each of mass $m$ and driven by a stochastic self-propulsion force $\f = f {\bf n}$ whose  direction ${\bf n} \equiv \(\cos \theta, \sin \theta\)$ undergoes
rotational diffusion, is given by,
\bea
m{\ddot{\mathbf{x}}}_i & = & -\gamma {\dot{\mathbf{x}}}_i +\sum_{i\neq j=1}^{N} \mathbf{f}_{ij} + f \mathbf{n}_i + \vartheta_i\,, \nn \\
\dot{\theta}_i & = & \xi_i\,.
\label{eq:abp}
\eea
where the $i$-th particle is subject to a friction $\gamma$ and a 
 thermal noise $\vartheta_i$ with zero mean and variance $2 k_B T \gamma \delta(t-t')$
that obeys fluctuation-dissipation relation. The rotational diffusion of the orientation of the propulsion force ${\theta}_i$ is described by an
athermal noise $\xi_i$, with  zero mean and correlation $\langle \xi_i(t) \xi_j(t') \rangle = 2 \tau_p^{-1} \delta_{ij} \delta(t-t')$. 
Its effect on the ${\mathbf{x}}_i$-dynamics is as an exponentially correlated vectorial noise with correlation time $\tau_p$,
which being unrelated to the drag $\gamma$, violates fluctuation-dissipation relation.
The inter-particle force $\mathbf{f}_{ij}$ is modelled via Lennard-Jones interaction 
with particle diameter $\sigma$. See~\cite{SI} for simulation details and units. 
Extreme dense active matter is  characterised by
large $\tau_p$, high P\'eclet number $f\sigma/k_BT$ and high densities. Here, we have fixed the
number density to be $1.2$, which is in the regime where this Kob-Andersen model~\cite{kob} 
of passive binary soft-spheres, shows dynamical arrest at low temperatures. Further, we focus on the strictly athermal
limit $T=0$; we have checked that our results hold when $T$ is small.

The set of equations (Eq.\,\ref{eq:abp}) for the assembly of particles are numerically integrated, using velocity Verlet algorithm, and 
we monitor the dynamics of  density relaxations and time series of energies, stresses etc., by changing $f$ at different values of $\tau_p$.
\\

\noindent
{\bf  Low persistence time: dynamical arrest}\\

\noindent
At high propulsion force, $f$, the material  is a fluid with 
time-correlations of density fluctuations measured via the self-overlap function $Q(t)$ relaxing diffusively 
(see Fig.\,1a(ii) and Fig.\,S2 in Supplementary Information~\cite{SI}). As $f$ is reduced, 
density relaxations slow down, until the onset of glass transition at $f=f_c(\tau_p)$, 
estimated by fitting the variation of the relaxation times $\tau_\alpha$ versus $f$  (Fig.\,1a(iii)) 
with a diverging power-law~\cite{kurchan}.
The glass transition boundary, obtained for a range of $\tau_p$ in this low persistence limit, can be fairly accurately described by an active generalisation of the well known 
Random First Order Transition (RFOT) theory~\cite{RFOT}, with an ``effective temperature'' that goes as $Af^2\tau_p/(1+B\tau_p)$ (see Fig.\,5),
$A$ and $B$ being fit parameters~\cite{ARFOT}. Indeed, recent studies~\cite{kurchan,berthier,ni,marchetti1,mandal1,flenner} 
are consistent with the predictions of this active RFOT theory, in the limit of low $\tau_p$.
This slowing down of particle motion, is also apparent in the time series of the mean kinetic energy 
$E(t)$ as one reduces $f$, i.e., the mean and variance of the kinetic energy decrease as $f \to f_c(\tau_p)$~(Fig.\,1a(i)). 
However, the mean kinetic energy appears to have a nontrivial scaling with $\tau_p$ (see Fig.\,S3 in Supplementary Information~\cite{SI}), over the range of $\tau_p$ values investigated. 
This deviation from equipartition, may be due to the fact that 
 the joint distribution of velocities and positions at steady state do not decouple~\cite{fodor2016}. 
\\


 \begin{figure*}[t]
{\includegraphics[scale=1.65]{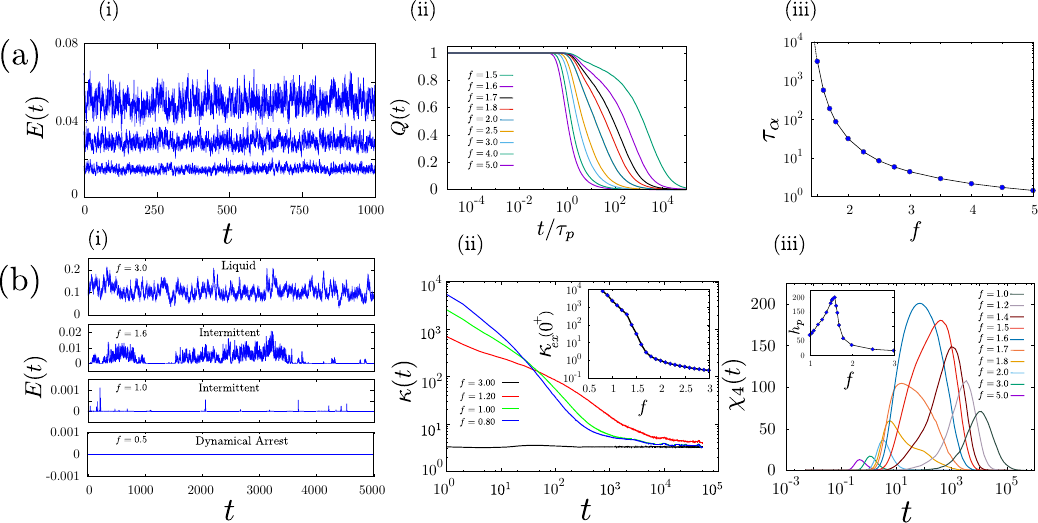}}
\caption{(color online). {\bf Dynamical arrest and Intermittency at low and intermediate $\tau_p$}. (a) {\it Low $\tau_p=1$}: 
(i) Kinetic energy time-series, $E(t)$,  at different 
$f=2.5, 2.0, 1.5$ ({\it top to bottom}), show regular fluctuations; both the mean and the variance reduce with decreasing $f$.
(ii) Density fluctuations, measured via the self-overlap function $Q(t)$, relax more slowly as the activity $f$ is reduced. 
(iii) The slowness of the relaxation dynamics is measured by the  
$\alpha$-relaxation time, $\tau_\alpha$, extracted from $Q(t)$, for each $f$.
The measured  $\tau_\alpha$ vs $f$ is fitted (solid line) using a diverging power-law form, which traces out the limit of 
dynamical arrest $f_c(\tau_p)$ for small $\tau_p$ in Fig\,5.
(b) {\it Intermediate $\tau_p=10^4$}: 
(i) Kinetic energy time series as $f$ is lowered, shows Gaussian fluctuations at high $f>f^*$, intermittent bursts and quiescence, and finally dynamical arrest when $f\leq f_c$.
(ii) Intermittency is characterised by the behaviour of the time dependent {\it kurtosis} of the kinetic energy 
time series, $\kappa(t)=\frac{\langle (E(t_0+t)-E(t_0))^4 \rangle}{{\langle (E(t_0+t)-E(t_0))^2\rangle}^2}$. We see that in the small $t$ end of this log-plot, 
 $\kappa(t)$ increases linearly as $t$ decreases and  should therefore diverge, when extrapolated to
$t\to 0$.  The dynamical order parameter is measured from the value of  $\kappa(t)$ at the earliest time that we can evaluate, i.e.
$t=0^{+}$. (Inset)  Variation of the dynamical order parameter, the {\it excess kurtosis}, $\kappa_{ex}(0^{+})$ with $f$. 
We use the point of inflection of this curve to determine the phase boundary to the intermittent phase. 
(iii)  The fluctuation $\chi_4(t) = \langle Q^2(t)\rangle - \langle Q(t)\rangle^2$, shows a peak at a time $t$ 
for different values of $f$. (inset) At a fixed $\tau_p=10^4$, the value of the peak height $h_p$ increases sharply as $f$ approaches $f^*(\tau_p)$ from above, 
then reduces again. The value of $f$ at which $h_p$ has a maximum, for different values of $\tau_p$, 
also marks the boundary between the liquid and the intermittent phase (see Fig.\,5).
} 
\label{f1}
\end{figure*}

\noindent
{\bf Intermediate persistence time: intermittent jamming and plastic yielding}\\

\noindent
At intermediate persistence times $\tau_p\gtrsim 10^3$, the relaxation dynamics is fundamentally different from that 
at low $\tau_p$. At high propulsion force, $f$, the particles move as a fluid as before,
with the time series of the mean kinetic energy showing typical Brownian fluctuations (Fig.\,1b(i)).
As $f$ is reduced, the local particle displacements start to show 
 spatial correlations, with a growing correlation length as one approaches $f=f^*(\tau_p)$
 (see Fig.\,S4 in Supplementary Information~\cite{SI}).
At and below the transition $f^*$, the average kinetic energy ($E(t)$)
shows sudden bursts (over a time interval $\tau_1$) with periods of quiescence or jamming
(over a time interval $\tau_2$), typical of  {\em intermittency}~\cite{frischbook}, as shown in Fig.\,1b(i), 
characterised by large fluctuations. 

To describe the dynamics of such statistical quantities that alternate between periods of quiescence 
and large changes over very short times, we monitor their time-dependent 
$4^{th}$-moment or {\it kurtosis} 
$\kappa(t)=\frac{[{\langle (E(t_0+t)-E(t_0))^4 \rangle}]}{[{{\langle (E(t_0+t)-E(t_0))^2\rangle}^2}]}$
\cite{frischbook,sachdeva}.
At $f>f^*(\tau_p)$, $\kappa(t)$ is nearly flat and close to $3$, 
indicating that the fluctuations are close to Gaussian. 
For $f\leq f^*(\tau_p)$, $\kappa(t)$ shows an increase at small $t$,  that becomes more pronounced with decreasing propulsion force $f$. 
We observe that $\kappa(t)$ shows a power law divergence at small $t$, a characteristic signature of intermittency~\cite{frischbook,sachdeva}.

This allows us to describe the intermittent phase by a dynamical order parameter, the {\it excess} kurtosis ${\kappa}_{ex}(0^{+})$,
of the increment in the kinetic energy over an infinitesimal time interval, 
which goes from $0$ (Gaussian distribution) when $f>f^*(\tau_p)$ to a finite value 
(indicative of broad non-Gaussian distributions) across a continuous transition at $f=f^*(\tau_p)$ (Fig.\,1b(ii)). 
The change in this order parameter is sharp for large values of  $\tau_p$ (inset of Fig.\,1b(ii)) and becomes more gradual 
as $\tau_p$ is reduced, indicating that at lower $\tau_p$, the transition from liquid to
intermittent phase  is more like a crossover. From the variation of this dynamical order parameter over the $\{\tau_p, f\}$ plane, we
 plot the non-equilibrium phase boundary in Fig.\,5.

Other quantities begin to show a broad distribution as $f\to f^*(\tau_p)$, such as in the time-correlation of density fluctuations, $Q(t)$. We see this
in the fluctuations, $\chi_4(t) = \langle Q^2(t)\rangle - \langle Q(t)\rangle^2$, a measure of the 
dynamical heterogeneity. At fixed $f$, $\chi_4(t)$ typically shows a peak at a time $t$ (which is less than $\tau_p$), as shown 
in  Fig.\,1b(iii), and we  denote the peak height as $h_p$ . The value of $h_p$ increases sharply as $f$ approaches $f^*(\tau_p)$ from above, 
then reduces again (see inset of Fig.\,1b(iii)). The value of $f$ at which $h_p$ has a  maximum, for different values of 
$\tau_p$ (see Fig.\,S5 in Supplementary Information~\cite{SI}),
provides another marker of the boundary between the liquid and the intermittent phase, indicated  by blue circles in Fig.\,5.

\begin{figure*}[t]
{\includegraphics[scale=1.2]{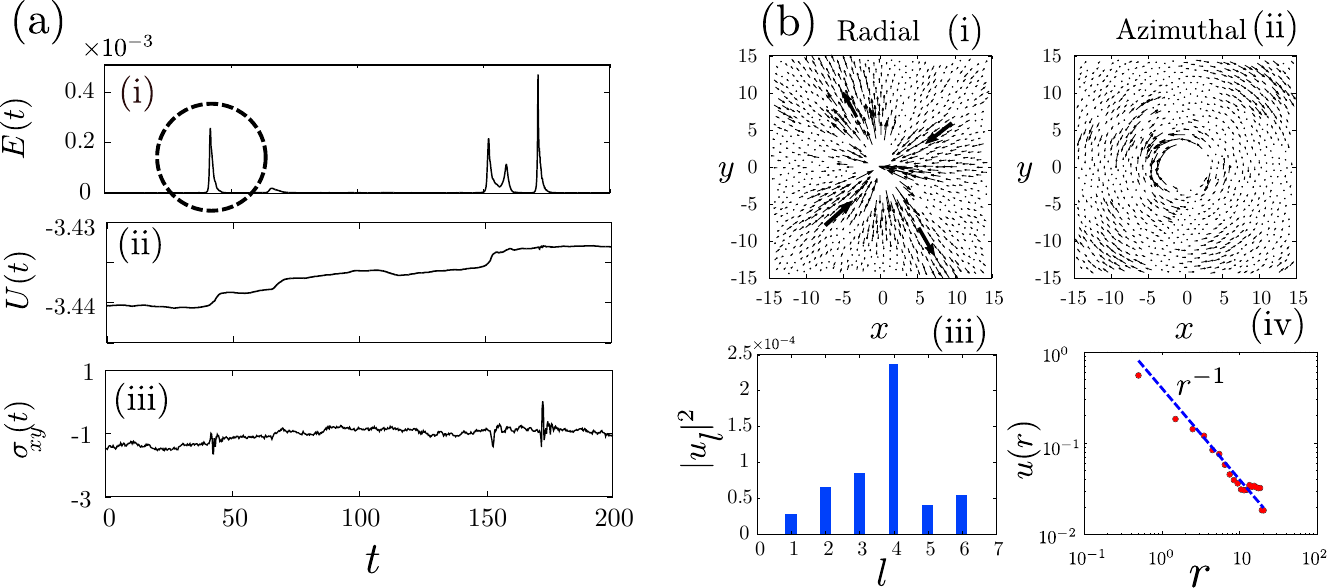}}
\caption{
{\bf Intermittent bursts associated with plastic yielding}. $\tau_p=10^4, f=1.0$ (marked by green triangle in Fig\,5).
(a) We monitor the time series of (i) mean kinetic energy $E(t)$, (ii) potential energy $U(t)$ and
(iii) local shear stress $\sigma_{xy}(t)$, at a value of $f< f^*(\tau_p)$ where the burst events are isolated and rare. Note that in the kinetic energy time series we have subtracted out the centre of mass contribution.
(b) The displacement field profiles surrounding a single kinetic energy burst event (encircled in (a)). (i) shows the radial component of the displacement field with a clear shear axis and quadrupolar symmetry.
(ii) shows the azimuthal component with a clear vorticity. (iii) the 4-fold symmetry shows up as a dominant $l=4$ mode in the power-spectrum of 
the radial component of the displacement field $u(r,\theta) = \sum_l u_l(r) \exp(- i l \theta)$, where we 
average over $r$ for better signal-to-noise.
(iv) Spatial profile of the radial component of the displacement shows a $1/r$ fall from the event.
This implies that the deformation associated with a single, isolated yield event is an Eshelby deformation.
}
\label{fig2}
\end{figure*}

Within the intermittent phase, we notice that the sudden increase
in kinetic energy during a burst is instantaneously accompanied by
a non-reversible release in the potential energy (Fig.\,\ref{fig2}a(ii),
as well as visible spikes in the local shear stress
(Fig.\,\ref{fig2}a(iii)). Thus, driven by persistent active stresses,
configurations of particles in the intermittent phase experience a
buildup of the elastic stress, a transient jamming ($E=0$), followed
by sudden yielding, seen as a burst of kinetic energy (Supplementary
Movies~\cite{SI}).

The bursts in kinetic energy are accompanied by local structural
reorganisations associated with sudden collective non-affine
displacements  of a {\it finite fraction} of particles (see Supplementary Movie 1~\cite{SI}).  These bursty features are
apparent in the thresholded displacements over a time $\tau$ (see
Fig.\,S6 in Supplementary Information~\cite{SI}), and more directly
in the spatial maps of the particle displacements.  This implies
that the intermittent steady state exhibits a continual  yielding
and jamming  of {\it macroscopically large structures}.  \\

\noindent
{\bf Plastic yielding: Eshelby deformations} \\

\noindent
Deep in the intermittent phase, these bursts in kinetic energy and associated plastic yielding, are rare and isolated (Fig.\,2a), 
allowing us to analyse the deformation field around a single burst event. 

The radial and azimuthal components of the displacement field ${\bf u}(r, \theta)$ surrounding the single yielding event,
 shows a quadripolar symmetry (Fig.\,2b(i)-(iii)) and a long range decay with radial distance that goes as $1/r$ (Fig.\,2b(iv)).
A similar feature is shown by the local elastic shear stress propagated as a consequence of a single yielding event. 
This is the well studied Eshelby deformation profile \cite{eshelby}, which describes elementary local deformations 
in an amorphous solid under external uniform shear \cite{argon, barratlemaitre}.
The unexpected appearance here of the Eshelby stress, is a result of local shear arising from internal stirring at the scale of the active particle.

As one moves towards the intermittent phase - liquid boundary from below, the bursts get more frequent and are bunched up. 
The distribution of the periods of intermittent bursts ($\tau_1$) and quiescence ($\tau_2$),
 is a power law with an exponential cut-off:
 $P(\tau_1)\sim \tau_1^{-\alpha} \exp(-\tau_1/\tau_{10})$ and $P(\tau_2)\sim \tau_2^{-\beta} \exp(-\tau_2/\tau_{20})$
(see Fig.\,3a for fit parameters),  with the cut-off moving to larger times as $f \to f^{*}(\tau_p)$ from below. 
 In the vicinity of $f^{*}(\tau_p)$, the distributions are power-laws, 
 $P(\tau_1)\sim \tau_1^{-2.6}$ and $P(\tau_2)\sim \tau_2^{-1.85}$ (Fig.\,3b). 

Each of these plastic events give rise to stresses that propagate through the material (see Supplementary Movies\,$2 \& 3$~\cite{SI}).
Outside the plastic zones, the rest of the material should respond elastically, with an anisotropic 
elastic kernel. However, the occurrence of multiple yielding events will result in 
strong correlations between events, one triggering another, that will make the kernel
isotropic, since the directions of local shear due to active forcing would be  randomly oriented.
\\

\noindent
{\bf Accumulated yielding,  turbulence}\\

\noindent
Since, for a small enough $f$, the plastic bursts are bunched up discrete events, the number of particles $n_c(\Delta t)$ 
that undergo irreversible displacement within a time window $\Delta t$ is the number 
of displaced particles per event times the number of events within the window $\Delta t$. We find that 
$n_c \sim N^2$ (Fig.\,3c), suggesting that each intermittent yielding event,
involves the collective displacement of a {\it finite fraction} of particles~\cite{puosi2015}. 
This implies that the occurrence of more and more such events will cause the material to 
flow at long time scales, with a time dependent viscosity $\eta(t)$, determined by the accumulated yielding upto  time $t$, 
eventually reaching a constant steady-state value for $t \gg \tau_p$.
This shows up, at the level of single tagged particle dynamics, 
as eventual diffusive motion when $t \gg \tau_p$ (see Fig.\,S7 and Fig.\,S9 in Supplementary Information~\cite{SI}),
and  coincides with the relaxation of the  self-overlap function $Q(t)$~\cite{SI}.

As we reduce the active force $f$, the tagged-particle diffusion coefficient decreases 
(Fig.\,S7 in Supplementary Information~\cite{SI}), and eventually vanishes as one approaches dynamical arrest.
As before, we obtain the dynamical arrest boundary $f=f_c(\tau_p)$ (see Fig.\,5), 
by fitting the data for the $\alpha-$relaxation time $\tau_\alpha$ measured from $Q(t)$ (Fig.\,S8 in Supplementary Information~\cite{SI}),
to a power-law divergence.

On the other hand, as we increase the forcing $f$ towards the phase boundary $f^{*}(\tau_p)$, we observe that the scale-free intermittency 
displays a kind of plastic turbulence~\cite{wensink,frey} in an actively stirred dense material. This is seen in the spatiotemporal dynamics 
of  the displacement fields which show large swirls; see Supplementary Movie 4~\cite{SI}, and in the power-spectrum of the kinetic energy density. 
The intermittent jamming-yielding due to local active stirring transfers energy from small scales and
dissipates it over larger scales, leading to an {\it inverse cascade}, where the energy spectrum crosses over from $E(k)\sim k^{-5}$ 
to $E(k) \sim k^{-5/3}$ at lower $k$ (Fig.\,3d). The crossover to the Kolmogorov spectrum happens at a scale corresponding to the scale of vorticity~\cite{tran}.
This stress production and dissipation gives rise to a non-equilibrium steady state with a finite energy-current.
\\


\begin{figure*}[h]
{\includegraphics[scale=1.3]{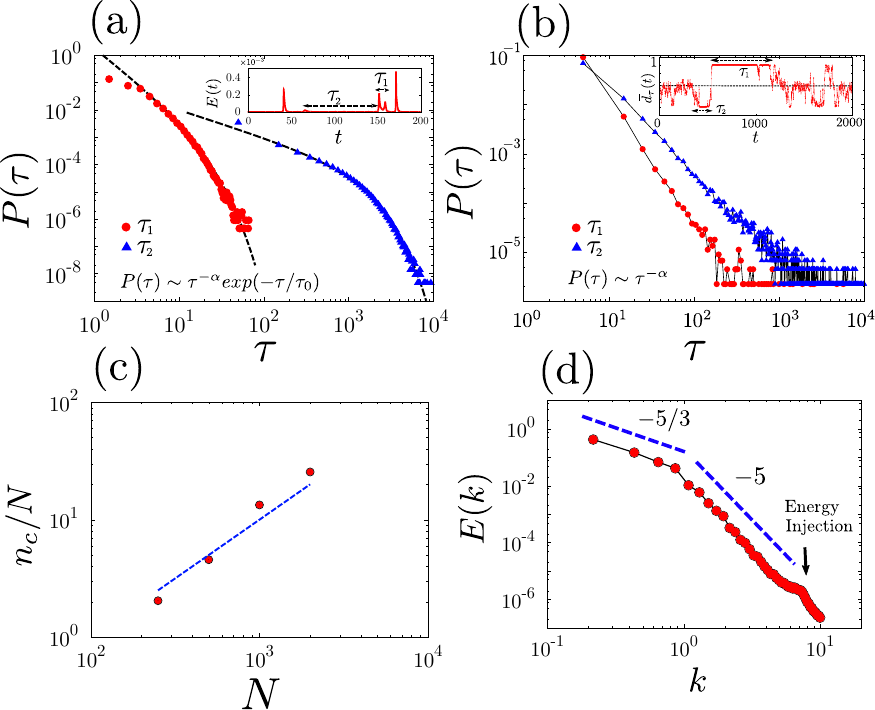}}
\caption{
{\bf Statistics of yielding, emergence of viscoelasticity and plastic turbulence.}
(a) Well within the intermittent phase ($f=1.0, \tau_p=10^4$), the distribution of the periods of intermittent bursts ($\tau_1$) 
and quiescence ($\tau_2$),  is a power law with an exponential cut-off, 
$P(\tau_1)\sim \tau_1^{-\alpha} \exp(-\tau_1/\tau_{10})$ and $P(\tau_2)\sim \tau_2^{-\beta} \exp(-\tau_2/\tau_{20})$, with fit values $\alpha=2.4$, $\tau_{10}=10$, $\beta=1$ and $\tau_{20}=875$.
We note that the cut-offs move to larger times as $f \to f^{*}(\tau_p)$ from below.
(b)  In the vicinity of the phase boundary $f^{*}(\tau_p)$, the distributions are power-laws, 
 $P(\tau_1)\sim \tau_1^{-2.6}$ and $P(\tau_2)\sim \tau_2^{-1.85}$. The data is shown for  $f=1.6, \tau_p=10^4$ (marked by red triangle in Fig.\,5).
(c) The intermittent yielding events involve the non-affine displacement of a finite fraction of particles, as seen in this plot of $n_c$, the  
number of particles that show a non-affine displacement within a {time window $\Delta t=10^4$}, versus total particle number $N$.
(d) The scale-free intermittency close to the phase boundary  $f^{*}(\tau_p)$, is associated with plastic turbulence as seen in 
the power-spectrum of the energy density $E(k)$  which shows an inverse cascade from an injection scale, shown by the arrow. 
Data shown for $f=1.6, \tau_p=10^4$. The crossover from a steep spectrum $k^{-5}$ to  the Kolmogorov spectrum $k^{-5/3}$ at lower $k$, is set by  the scale of the vorticity.}
\label{fig3}
\end{figure*}


\noindent
{\bf Infinite persistence: jamming/unjamming}\\

\noindent
Analysis of the $\tau_p=\infty$ limit, brings in a new facet of extreme active matter.
This limit corresponds to a situation where the 
initial directions of particle self-propulsion are {\it quenched in random directions}. From being a fluid with mobile particles
at large $f$, the assembly jams at $f^*(\infty) = 1.67$, where the kinetic energy goes to zero as $\sim \vert f - f^*(\infty)\vert^{3/2}$  (Fig.\,4a,b). 
The distribution of  forces $P({\bf F})$ changes from a broad distribution with exponential tails to a delta-function at ${\bf F}=0$ at $f^*(\infty)$; the jamming
transition is associated with a force-balanced configuration of the soft particles (Fig.\,4c).
The width of $P({\bf F})$ and the mean kinetic energy
go continuously  to zero as $f$ approaches $f^*(\infty)$ (inset Fig.\,4c). This allows us to  identify $f^*(\infty)$ 
as a jamming/unjamming force threshold for active yielding. 
As discussed in a recent work~\cite{liao&xu2018}, the density-dependent $f^*(\rho, \infty)$ will trace
out a yielding line in the jamming phase diagram~\cite{liunagel} of dense amorphous materials, 
with active forcing being the control variable. Thus, we expect to find critical behaviour in the proximity of $f^*(\infty)$.

In the vicinity of $f^*(\infty)$, we map the contact network of particles and evaluate the total
force between pairs of soft particles in contact; we find that these forces are distributed along force-chains (Fig.\,4d(iii)).
With increasing $f$, the force chains dynamically remodel, as the structures relax; see Fig.\,4d(i),(ii).
Likewise, dynamical reorganisation of the force network also occurs when we move slightly away from the jammed regime,
by decreasing $\tau_p$, keeping $f$ fixed. Under these conditions, 
the dynamics of force-chains typically show periods of jamming in a force-balanced configuration, interspersed with bursts of remodelling.
As one approaches the jamming threshold at $\tau_p=\infty$, the mean lifetime of the force-balanced configurations diverges  as
$\tau_{{\footnotesize F}} \sim \tau_p^{z}$, with a new dynamical critical exponent $z\approx 0.71$ (Fig.\,4e).
\\

  
 \begin{figure*}[h]
{\includegraphics[scale=0.45]{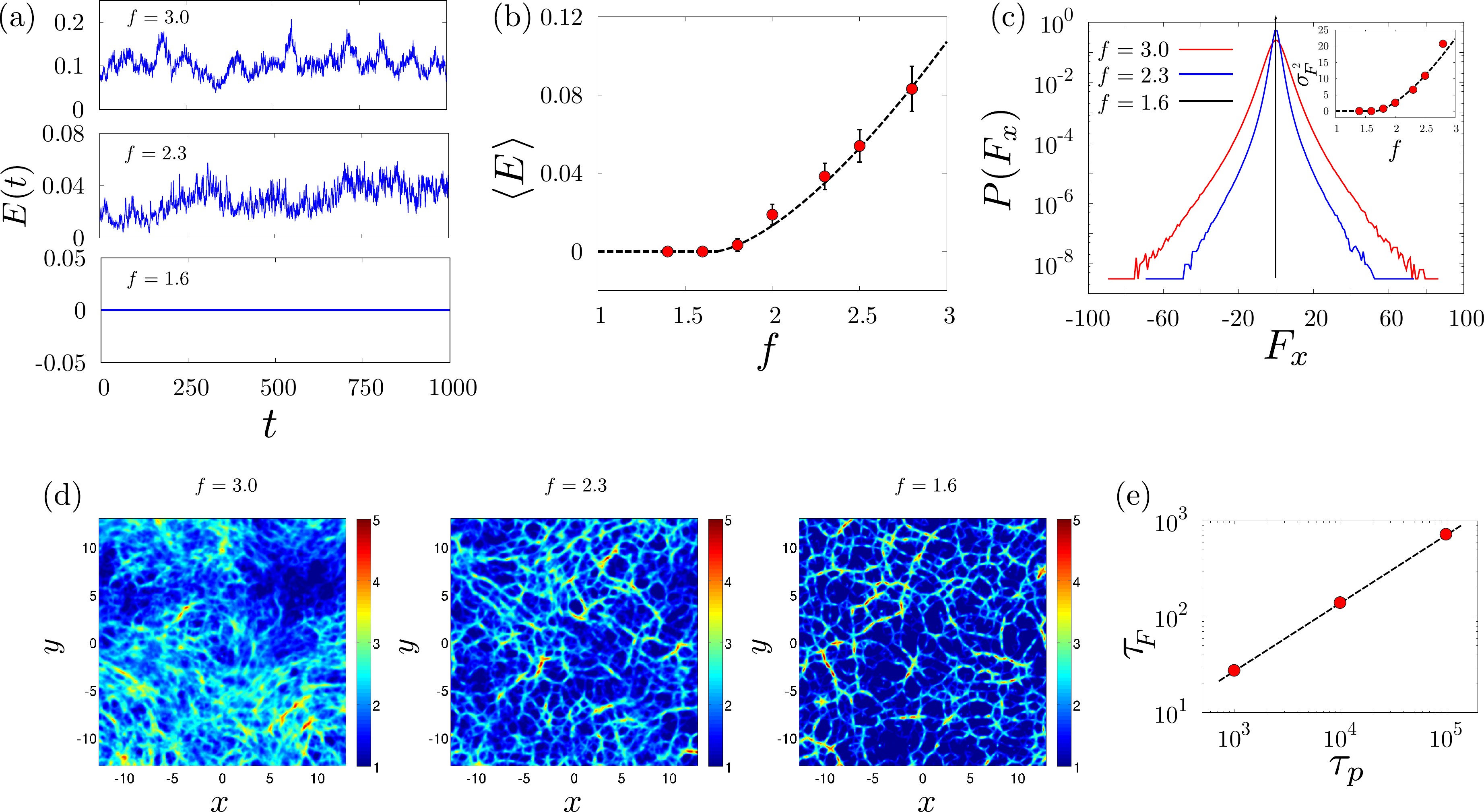}}
\caption{(color online). {\bf Jamming at infinite $\tau_p$ and force-chains}. (a) Kinetic energy time series as $f$ is lowered, 
shows complete jamming at $f\approx 1.6$. (b) The variation of the mean kinetic energy with $f$, shows a continuous transition at $f=f^*(\infty)= 1.6$, that goes
as $\langle E\rangle \sim \vert f - f^*(\infty)\vert^{3/2}$, shown with dashed line. (c) The probability distribution of the $x$-component of the total force acting on a particle,
$P(F_x)$ at different values of $f$, is broad with exponential tails, with a width that decreases continuously with $f$ (inset). 
At the jamming transition, $f=f_c$, the distribution becomes a delta-function at $F_x=0$, the force-balanced state.
(d) At the jamming critical point, the forces on the particles are distributed along force-chains, as
highlighted in the colour map of thresholded forces. Away from the jamming critical point, either by increasing $f$ keeping $\tau_p$ fixed, or 
decreasing $\tau_p$ keeping $f$ fixed, the force-chains dynamically remodel whilst still being embedded in a static contact network. 
This is seen as a blurring of the colour map of thresholded forces (at $f=3.0$ and $2.3$) away from the sharp force-chains at $f=1.6$.
(e) The dynamics of the force-chains show a distribution of lifetimes in the force-balanced configuration.
The  mean lifetime of the force-balanced configurations, computed for $f=1.6$ at varying $\tau_p$, diverges as one
moves towards the jamming critical point as a power-law,  $\tau_{{\footnotesize F}} \sim \tau_p^{z}$, with $z=0.71$ (shown with dashed line).
}
\end{figure*}


\noindent
{\bf Discussion}\\

\noindent
Putting all this  together our study suggests a rather rich phase
diagram (Fig.\,5).  The study of extreme dense active matter allows
us to explore the crossover between glass physics, where the dynamics
proceeds by density relaxation, and jamming-yielding physics, where
the dynamics is controlled by stress  buildup and release via
macroscopic flows. We emphasize that the intermittent plastic
deformation and turbulent flows are constitutive and not response
to an externally imposed stress.

While our present study was done at $T=0$, we have checked that
including a small temperature via a thermal noise $\vartheta_i$
gives similar results, as long as the P\'eclet number is high; the
crossover behaviour from these different regimes are likely to be
quite subtle.

We have fixed the overall density of particles in our study,
however~\cite{liao&xu2018} shows that the athermal jamming transition
at  $\tau_p=\infty$ occurs over a range of densities, making this
active jamming critical point, density dependent. If however we
make the density very low, while still keeping $\tau_p=\infty$, we
would arrive at a jammed gas phase, with isolated islands of jammed
material~\cite{Bulbul}.

Are there natural or synthetic realisations of extreme active matter?
Herds of animals, such as penguins or bulls, dense collection of
vehicles, ants or microbots and even trite examples such as a herd
of rugby players, could be possible realisations.  Promising
candidates for extreme active matter are monolayers of persistently
motile cells; indeed~\cite{garcia2015} observe jamming-yielding
behaviour in such monolayers of epithelial cells.  It would be
challenging however to construct synthetic realisations of extreme
active matter, and we eagerly look forward to controlled experimental
studies on these.\\\\

 \begin{figure*}[h]
{\includegraphics[scale=1.5]{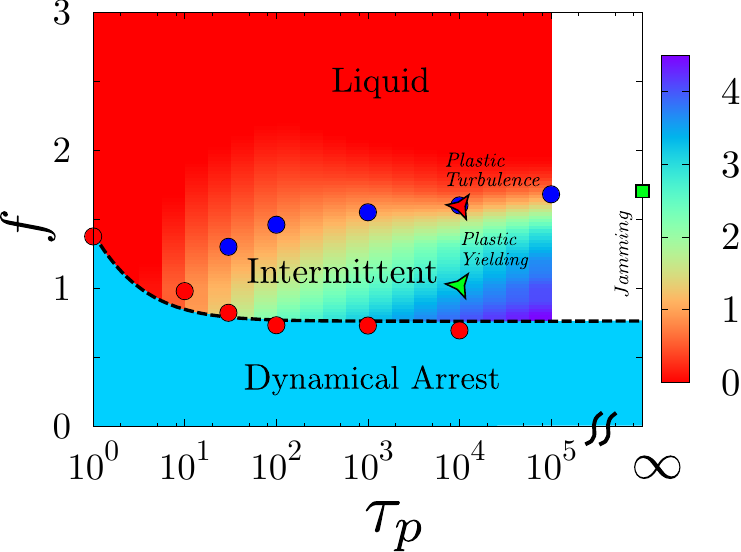}}
\caption{(color online). {\bf Dynamical phase diagram in $f-\tau_p$ at fixed density}.
At low $\tau_p$ ($0 < \tau_p < 10$), there is a direct transition at $f = f_c$ (red dots), from the liquid to a dynamically arrested phase, as 
measured from the the divergence of $\tau_\alpha$, the slow relaxation of the density fluctuations (Fig.\,1a(i)). This transition to the 
dynamically arrested state continues into larger values of $\tau_p$ (red dots).
The dashed line is a fit to the active RFOT theory~\cite{ARFOT} with an effective temperature $Af^2 \tau_p/(1+B \tau_p)$, where $A$ and $B$ are fit parameters.
At larger values of $\tau_p > 10$, a new {\it intermittent} phase, intervenes between the liquid and the dynamically arrested phase.
The liquid-intermittent phase boundary is obtained from the dynamical order parameter, the excess kurtosis $\kappa_{ex}$ (Fig.\,1b(ii))
and the peak height $h_p$ (Fig.\,1b(iii)), shown with blue circles. 
This transition is very sharp and continuous at large values of $\tau_p$, and gets progressively broader and less defined at lower values
of $\tau_p$ (colour bar represents the value of $\kappa_{ex}(0^{+})$ in $\log_{10}$ for different $f$ and $\tau_p$). This intermittent phase
is characterised by bursts of plastic yielding and plastic turbulence close to the transition to the liquid phase.
At $\tau_p = \infty$, the assembly shows a sudden transition from a liquid to a jammed configuration at a force threshold,
where the particles suddenly get into a force-balanced configuration (Fig.\,4d).
}
\end{figure*}

\noindent
{\bf {\em Acknowledgements}}\\
\noindent
We would like to thank S. Sastry, B. Chakraborty, K. Ramola, L. Berthier, J.-L. Barrat and M. Muthukumar for useful discussions, and S. S. Ray
for help in analysing aspects related to turbulence. We thank IISc and NCBS for computing facilities. R.M. acknowledges  
funding from Simons Centre for the Study of Living Machines at NCBS through the Simons Career Development Post-Doctoral Fellowship.
P. C. acknowledges financial support from CEFIPRA Grant No. 5604-1.


\clearpage

\renewcommand{\thefigure}{S\arabic{figure}}
\setcounter{figure}{0}    

\section{Supplementary Information}

\subsection{Model and Methods} \label{ap:modmeth}

For our study, we consider a well-studied two-dimensional model, {\it viz.}, a 65:35 binary Lennard-Jones (LJ) mixture~\cite{kob}, with particles interacting 
via the following potential,
\begin{equation}
 V_{ij}(r)=4 \epsilon_{\alpha \beta} \[\(\frac{\sigma_{\alpha \beta}}{r}\)^{12}-\(\frac{\sigma_{\alpha \beta}}{r}\)^{6}\]
\end{equation} 
where $r$ is the distance between the $i$-th and the $j$-th
particle, {\textit{i.e.}}, $r=|\mathbf{r}_i-\mathbf{r}_j|$ and
$\alpha$, $\beta$ represent either $A$-type or $B$-type particles. 
The strength and the range of the interaction are set by $\epsilon_{\alpha \beta}$ and $\sigma_{\alpha \beta}$ respectively.
In our simulation we have chosen the values of $\sigma_{\alpha
\beta}$ and $\epsilon_{\alpha \beta}$ to be: $\sigma_{AB}=0.8
\sigma_{AA}$, $\sigma_{BB}=0.88 \sigma_{AA}$, $\epsilon_{AB}= 1.5
\epsilon_{AA}$, $\epsilon_{BB}=0.5 \epsilon_{AA}$. The composition of the $A:B$ mixture helps to avoid crystallisation, in the absence of activity. The potential
has been truncated at $r^{c}_{\alpha \beta}=2.5 \sigma_{\alpha
\beta}$ and has been shifted accordingly such that
both the potential and the force remain continuous at the cut-off. The
unit of length and energy in our simulation are set by  $\sigma_{AA}=1$
and $\epsilon_{AA}=1$ and the study is done for an  overall number
density of  $\rho=1.2$. All the particles have the same mass ($m=1$)
and the time unit is $\tau_{LJ} \equiv \sqrt{m{\sigma_{AA}^2}/\epsilon_{AA}} = 1$.

\begin{figure}[h]
\centerline{\includegraphics[scale=0.4]{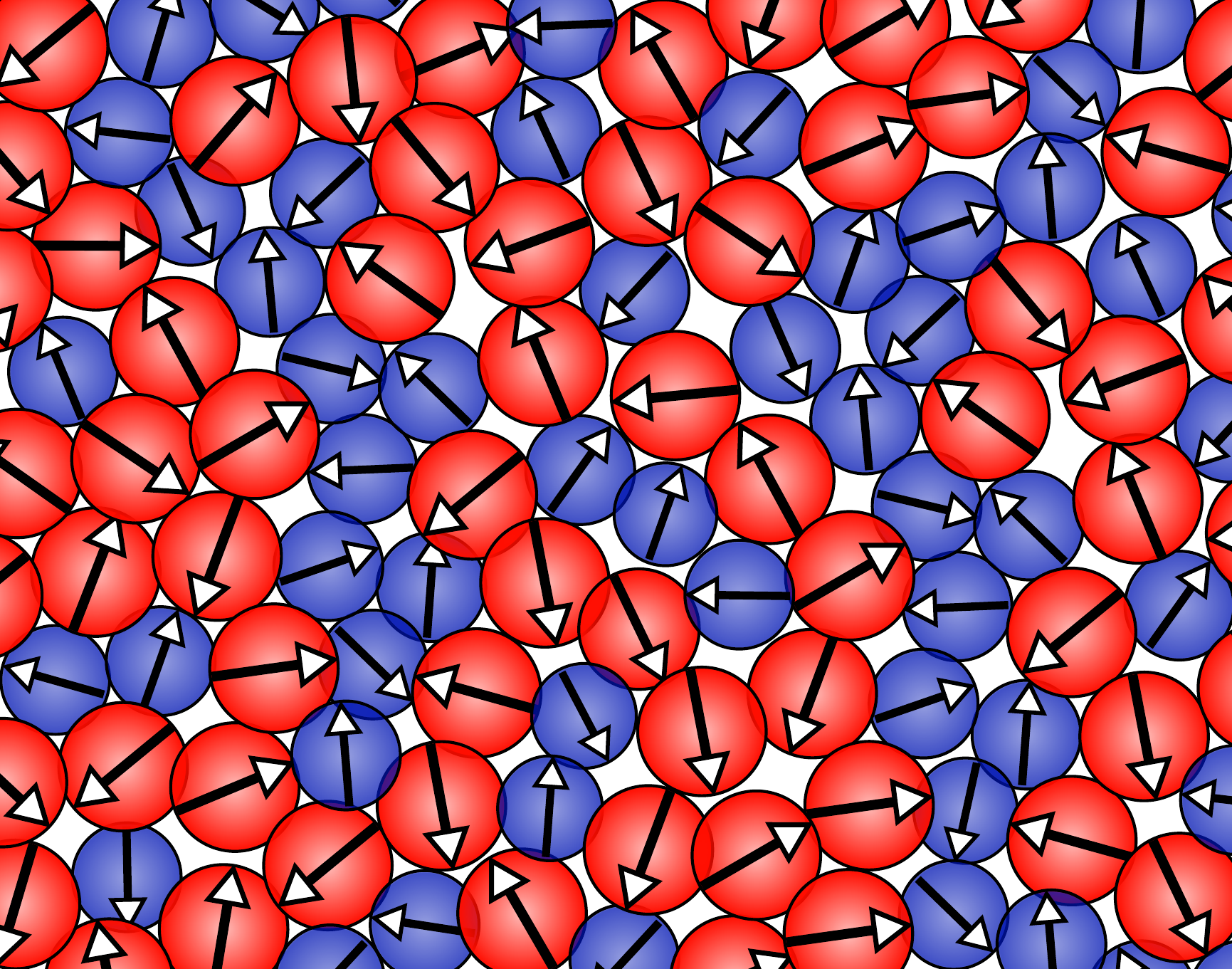}}
\caption{Schematic of a dense assembly of bi-disperse $A$(red)-$B$(blue) soft Lennard-Jones particles, with attached arrows denoting the direction of 
self-propulsion.}
\label{fs0}
\end{figure}

We study the assembly under athermal conditions, with  a self propulsion 
force of magnitude $f$ along a unit vector associated with each particle, in the presence 
of a Langevin bath. Thus, the equation of motion for each particle is,
\begin{equation}
m{\dot{\mathbf{v}}}_i=-\gamma \mathbf{v}_i +\sum_{j=1, j\neq i}^{N} \mathbf{f}_{ij} + f \mathbf{n}_i
\end{equation}
where $m$ is the mass of a particle, $\mathbf{v}_i$ is the velocity of the $i$-th particle, $\gamma$ is the friction coefficient. For our study, we choose $\gamma=1$.
Thus, the particle inertia relaxation time $\tau_\gamma \equiv \frac{m}{\gamma}$ is comparable to $\tau_{LJ}$.
$\mathbf{f}_{ij}$ is the LJ interaction force between the $i$ and $j$-th particle, 
$\mathbf{n}_i$ is the unit vector associated with the $i$-th particle along which the propulsion force is being imparted, $f$ is the strength of the propulsion force. See Fig.\,\ref{fs0}
for a schematic snapshot of a typical configuration, with the propulsion direction of each particle being indicated by an arrow. The dynamics of the direction of $\mathbf{n}_i$ follows simple rotational diffusion equation with diffusion constant $D_R \propto \tau_p^{-1}$ ($\tau_p$ is the persistence time). In our study, we will  tune both $f$, and $\tau_p$, and study the dynamical behaviour of the assembly of the active particles. The number of particle used in the simulation varies between $N=1000-10000$. All data presented here have also been averaged over $32-96$ independent realisations, if not mentioned otherwise.

\medskip

\noindent

To characterise the dynamics of the dense liquid, under steady state or transient conditions, as we vary the active forcing ($f$), for various choice of $\tau_p$, we  measure the mean squared displacement ($MSD$)  ${\Delta}^2(t)$ and self-part of the two-point overlap correlation function $Q(t)$, defined as,

\begin{equation}
{\Delta}^2(t)=\langle \frac{1}{N} \sum_{i} \mid{\bf r}_i(t_0)- {\bf r}_i(t+t_0)\mid^2\rangle
\end{equation}

\begin{equation}
Q(t) =\langle \frac{1}{N} \sum_{i} q(\mid{\bf r}_i(t_0)- {\bf r}_i(t+t_0)\mid)\rangle
\end{equation}
where,
\begin{equation}
q(r)=
\left\{
        \begin{array}{ll}
                1  & \mbox{if } r \leq b\\
                0  & \mbox{otherwise}
        \end{array}
\right.
\end{equation}
 $\langle \cdots \rangle$ represents an average over the time origin $t_0$, $N$ is the number of particles in the system and the parameter $b$ is associated with the typical vibrational amplitude of the caged particles. Throughout our analysis we have used $b=0.3$ and we have verified that our results are insensitive to moderate changes in $b$. 
 
\begin{figure}[h]
{\includegraphics[width=7.5cm]{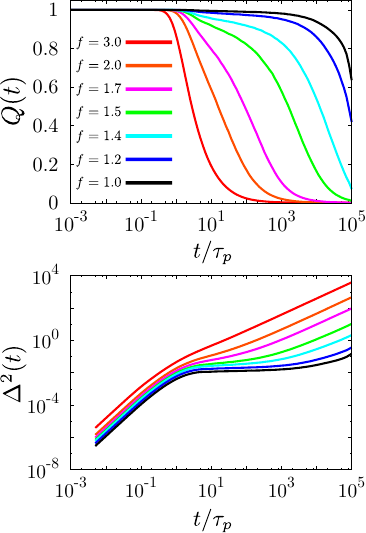}}
\caption{$\tau_p=1$. (Top) Self-overlap function, $Q(t)$, for different values of active forcing $f$, as indicated. (Bottom) Corresponding mean squared displacement, $\Delta^{2}(t)$.
Both quantities show that relaxation timescales increase with decreasing $f$.}
\label{fs1}
\end{figure}

Plots showing the variation of $\Delta^2(t)$ and $Q(t)$, with $f$ and $\tau_p$ are shown in Fig.\,\ref{fs1}, \ref{fs3}, \ref{fs6}.
 
Another quantity of interest is the displacement field, calculated over a time scale of $\tau_\alpha$; see Fig.\,\ref{fs2}. Using that, we calculate the two point spatial correlation function, defined as
$C(r)=\langle \vec{x}_{\tau_\alpha}(\vec{r})\cdot \vec{x}_{\tau_\alpha}(\vec{0}) \rangle$ where $\vec{x}_{\tau_\alpha}(\vec{r})$ is the displacement of the particles at position $\vec{r}$ over the timescale $\tau_\alpha$. To extract the lengthscale ($\zeta$) associated with the size of this cooperatively rearranging regions we use the relation $C(\zeta)=1/e$, and we monitor how $\zeta$ varies with $f$; see Fig.\,\ref{fs2}.

\newpage

\subsection{Supplementary Plots}

\begin{figure}[h]
{\includegraphics[width=7.5cm]{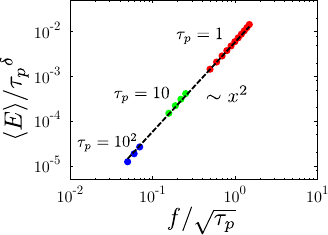}}
\caption{Dependence of mean kinetic energy on $f$ and $\tau_p$, shown here as a scaling plot,
$\langle E \rangle \propto \tau_p^{\delta}\,G(f \tau_p^{-\alpha})$, where $\delta=0.11$, $\alpha=0.5$,  for values of $1 \leq \tau_p \leq 100$. For smaller values of $\tau_p$, one might expect a crossover.}
\label{fs1a}
\end{figure}

\begin{figure}[h]
{\includegraphics[width=7.5cm]{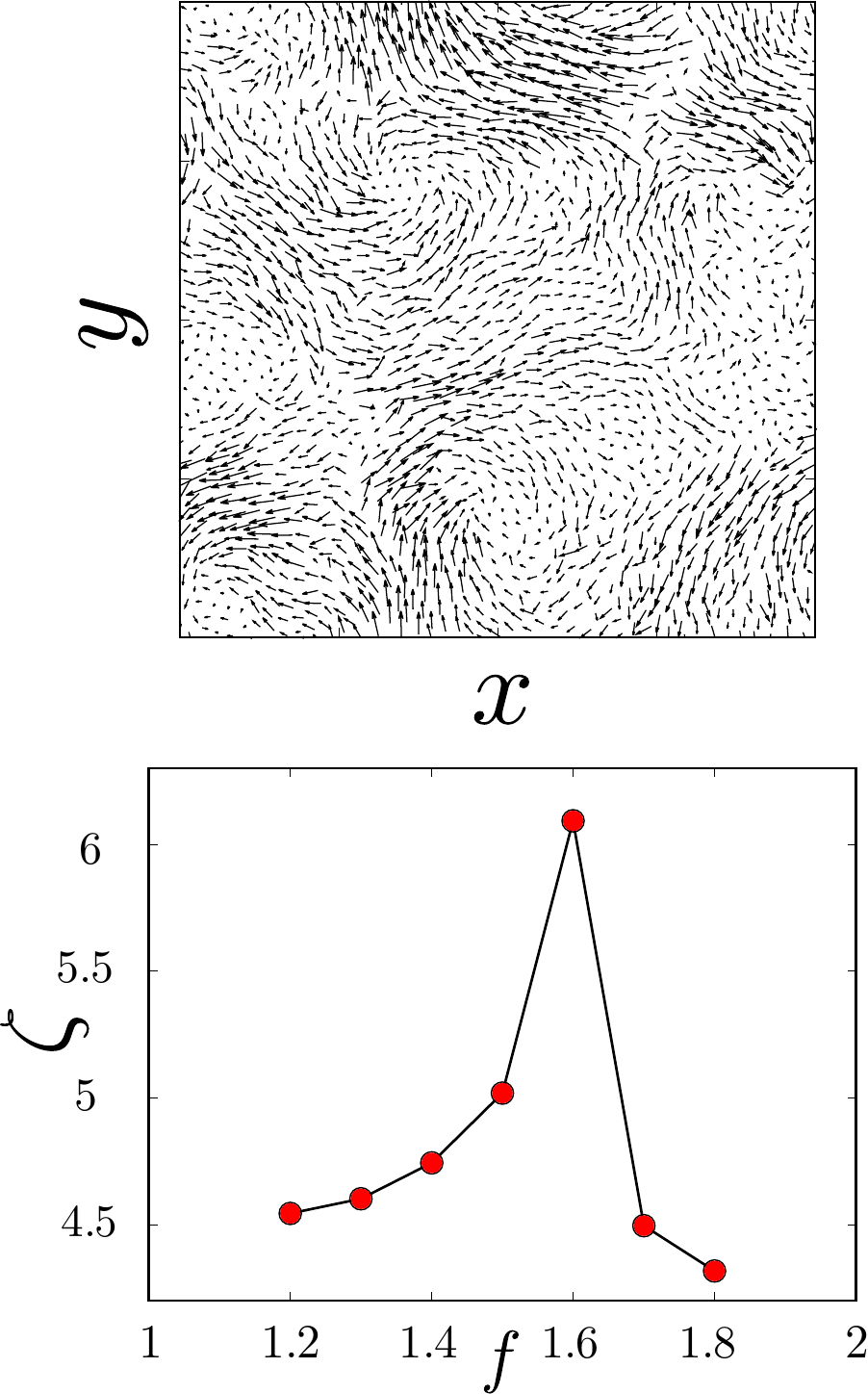}}
\caption{$\tau_p=10^4$.  (Top) A typical displacement field map (calculated over timescale $\tau_\alpha$) at $f=1.4$, showing strong spatial correlations, and the emergence of swirl-like collective motion during $\alpha-$relaxation time scale. (Bottom) Variation of dynamical length scale, $\zeta$, calculated from spatial correlation function $C(r)$, defined in the SI text, 
as the active forcing ($f$) is varied; it peaks at the transition region ($f\sim 1.6$) between the transition from the intermittent regime to the liquid-like regime.}
\label{fs2}
\end{figure}

\begin{figure}[h]
{\includegraphics[width=7.5cm]{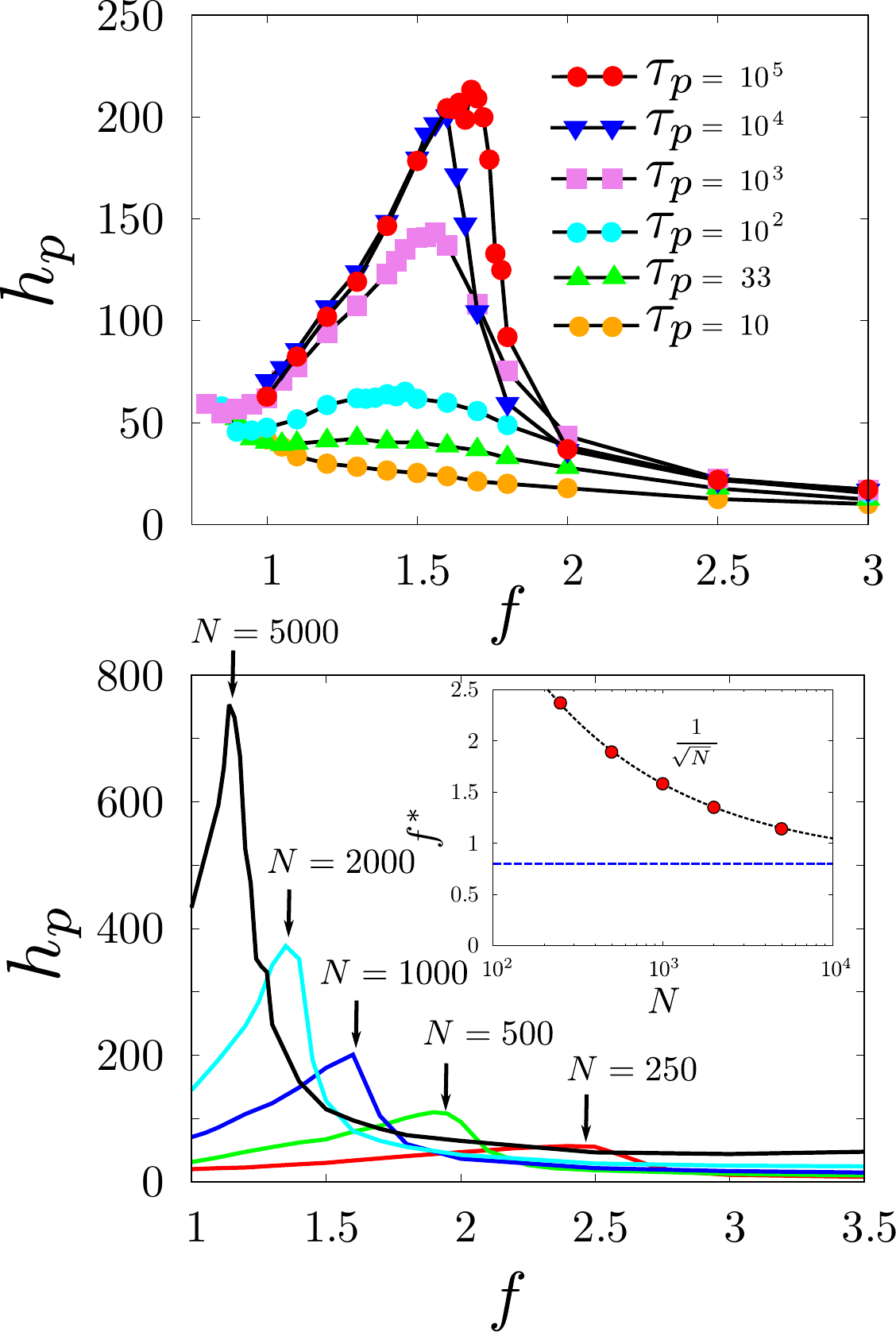}}
\caption{(Top) Variation of peak height ($h_p$) of $\chi_4(t)$, the fluctuation of the overlap function $Q(t)$, with active forcing $f$, for various values of $\tau_p$, shows non-monotonic behaviour in the range $\tau_p > 10$. The locus of $f$ at which $h_{p}$ has a maximum, for each $\tau_p$, defines the boundary between intermittent and fluid regimes. (Bottom) The variation of  $h_p$ with $f$, for various system sizes, shows that the fluctuations increase with increasing system size $N$, indicating an underlying dynamical transition. The inset shows that $f^{*}$, the force at which the peak occurs, remains finite in the thermodynamic limit.}
\label{fs5}
\end{figure}

\begin{figure}[h]
{\includegraphics[width=8cm]{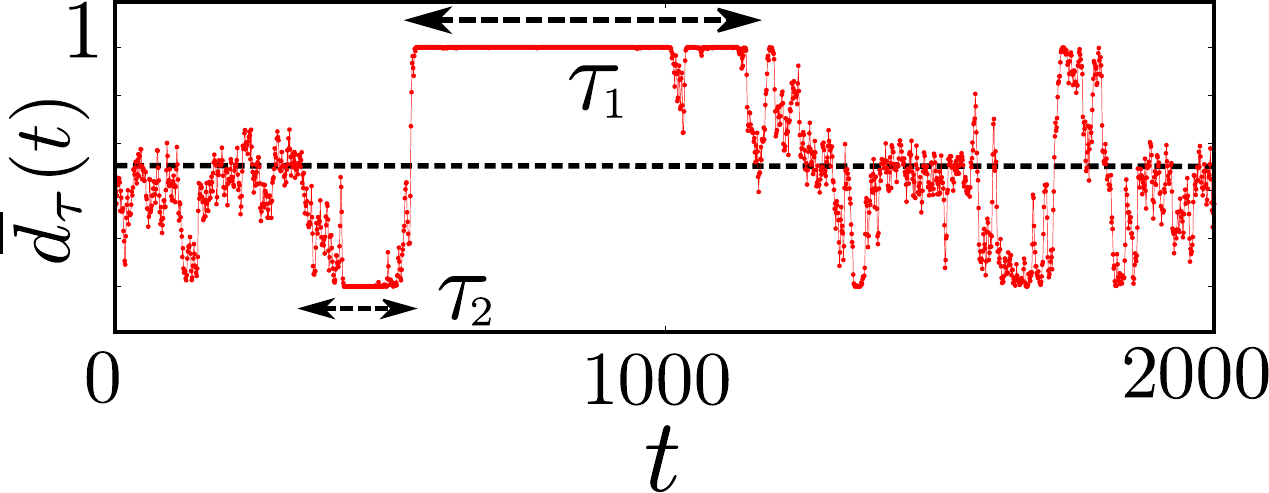}}
\caption{$\tau_p=10^4$. In the intermittent phase (and in the vicinity of the liquid-intermittent boundary), the system 
switches between a jammed and flowing region, as captured by
the displacement overlap function $\overline{d}_{\tau}(t)$, defined as 
$d_{\tau}(t)=1$ if the displacement between time $t-\tau$ and $t$ is more than $a$ and $d_{\tau}(t)=0$ if it is smaller than $a$, where we chose $a=0.1$ and $\tau=1.0$. Here, the propulsion force, $f\sim1.6$.
}
\end{figure}

\begin{figure}[h]
{\includegraphics[width=7.5cm]{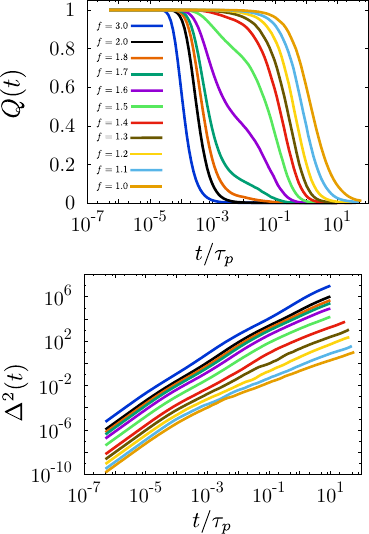}}
\caption{$\tau_p=10^4$. (Top) Self-overlap function, $Q(t)$, for different values of active forcing $f$, as indicated. (Bottom) Corresponding mean squared displacement, $\Delta^{2}(t)$. The relaxation functions show a change in behaviour around $f=1.6$,
which is where  the relaxation timescale $\tau_\alpha$ shows a jump (see Fig.\ref{fs4}) and the peak value of 
$\chi_4(t)$ shows a maximum, with changing $f$, as shown in Fig.\ref{fs5}.}
\label{fs3}
\end{figure}

\begin{figure}[h]
{\includegraphics[width=7.5cm]{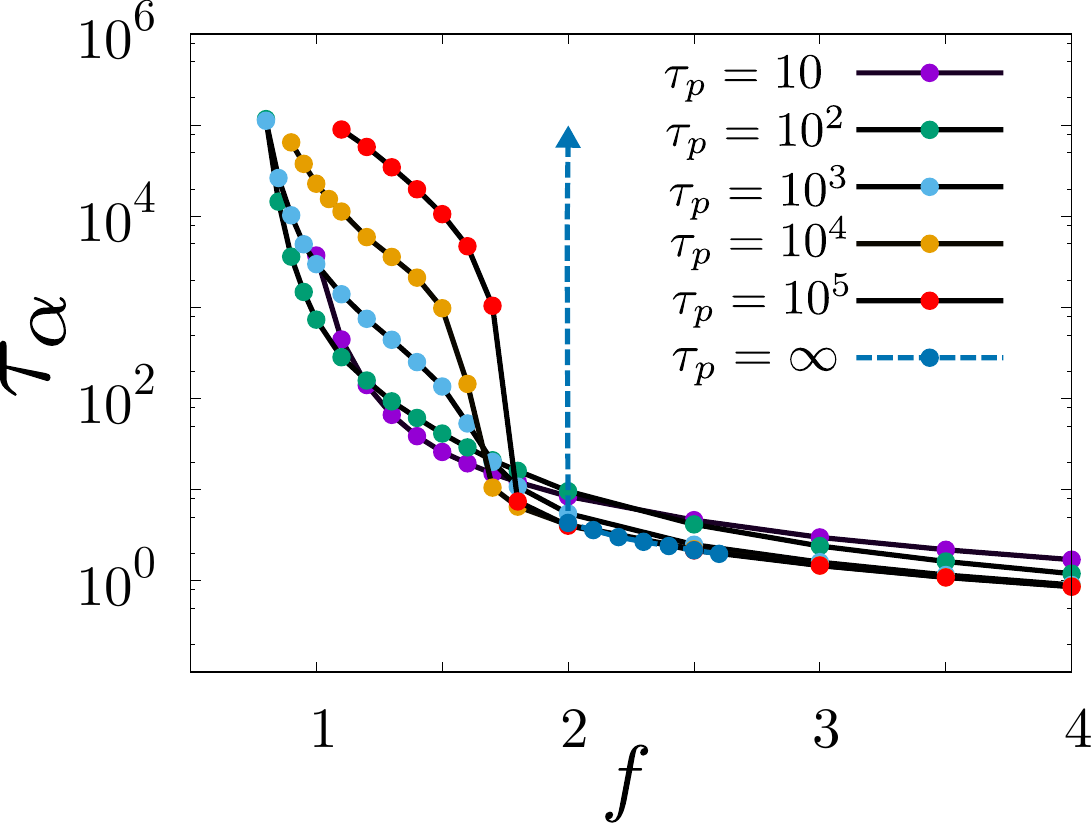}}
\caption{Relaxation timescale $\tau_{\alpha}$ extracted from $Q(\tau_\alpha)=1/e$, 
as a function of $f$ for a range $\tau_p$ (see labels).
For $\tau_p > 10$, $\tau_{\alpha}$ vs $f$ has a jump, with the location 
corresponding to where there is a peak in $h_{\chi_4}$, as shown in Fig.\ref{fs5}. }
\label{fs4}
\end{figure}

\begin{figure}[h]
{\includegraphics[width=7.5cm]{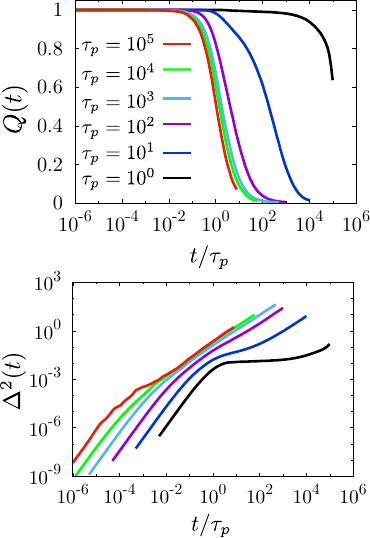}}
\caption{$f=1$. (Top) Self-overlap function, $Q(t)$, for different values of persistence time $\tau_p$ of self-propulsion, 
as indicated. (Bottom) Corresponding mean squared displacement, $\Delta^{2}(t)$. 
For $\tau_p \geq 10^3$,  the characteristic relaxation timescale is $t/\tau_p \approx 1$,
and diffusive motion is also seen to set in, beyond this timescale.}
\label{fs6}
\end{figure}

\clearpage

\subsection{Supplementary Movies}

\begin{enumerate}
\item Movie 1 ({\em 2intermittency.avi}): The intermittent bursts in the time series of the mean kinetic energy (below), and corresponding thresholded displacement field showing localised events. Here, the self-propulsion force $f=1$, and the persistence time $\tau_p=10^4$.
\item Movies 2 $\&$ 3 ({\it 4compress.avi $\&$ 4shear.avi}): For $\tau_p=10^4$ and
$f=1$, we observe propagation of both compressive and shear stresses, following a structural rearrangement event, marked by a spike in the kinetic energy.
\item Movie 4 ({\it 5swirl.avi}): For $\tau_p=10^4, f=1.4$, we observe swirling patterns in the displacement field, interspersed with periods of quiescence. The corresponding  time series of kinetic energy (shown below) shows bursts of activity corresponding to the periods of swirl motion.

\end{enumerate}

\end{document}